\documentstyle[preprint,aps,version2]{revtex}
\begin{document}
\draft
\preprint{\vbox{Submitted to Physical Review {\bf C}\hfill
	UC/NPL-1108}}
%
%
\tolerance=10000
\hbadness=10000
\tighten
\def\partder#1#2{{\partial #1\over\partial #2}}
\def\pslash{p\llap{/}}
\def\llpap#1{\hbox to 0pt{\hskip-.6em#1}}
\def\wp{p\llpap{\lower.1em\hbox{\_}}}
\def\qs{{\bf q}}
\def\bfy{{\bf y}}
\def\bfq{{\bf q}}
\def\bfp{{\bf p}}
\def\bfP{{\bf P}}
\def\bfsigma{{\bf \sigma}}

\begin{title}
Comparison of $K^+$ and $e^-$ Quasielastic Scattering 
\end{title}
\author{J.~Piekarewicz}
\begin{instit}
Supercomputer Computations Research Institute,
Florida State University, Tallahassee FL 32306
\end{instit}
\author{J.~R.~Shepard}
\begin{instit}Department of Physics, University of Colorado, 
Boulder, CO 80309-0446
\end{instit}
\begin{abstract}
	We formulate $K^+$-nucleus quasielastic scattering in a manner 
which closely parallels standard treatments of $e^-$-nucleus quasielastic 
scattering.  For $K^+$ scattering, new responses involving scalar 
contributions appear in addition to the Coulomb (or longitudinal) and
transverse $(e,e')$ responses which are of vector character.  
We compute these responses using both nuclear matter and finite nucleus
versions of the Relativistic Hartree
Approximation to Quantum Hadrodynamics including RPA correlations.
Overall agreement with measured $(e,e')$ responses and new $K^+$
quasielastic scattering data for $^{40}$Ca at $|\qs|=500$ MeV/c is good.
Strong RPA quenching is essential for agreement with the Coulomb response.
This quenching is notably less for the $K^+$ cross section even though the
new scalar contributions are even more strongly quenched than the
vector contributions.  We show that this ``differential quenching''
alters sensitive cancellations in the expression for the $K^+$ cross section 
so that it is reduced much less than the individual responses.  We 
emphasize the role of the purely relativistic distinction between vector
and scalar contributions in obtaining an accurate and consistent 
description of the $(e,e')$ and $K^+$ data within the framework of
our nuclear structure model.
\end{abstract}
%
\newpage
\narrowtext
%
\section{INTRODUCTION}
 
	Quasielastic electron scattering from nuclei has long been
used to study various aspects of nuclear structure.  The strong
quenching of the measured Coulomb response relative to Fermi gas 
estimates is one prominent example\cite{electdat,electdatb,coulthy}.  
Such studies are
facilitated by (1) the (reasonably) well-understood nature of the
fundamental $eN$ interaction and (2) by the relative weakness 
of that interaction.  However, this same interaction suffers from some
important shortcomings in that it probes the nucleus in a very 
restricted fashion.  Hence only two independent observables exist
(without making polarization measurements), namely the Coulomb 
(or longitudinal) and
transverse responses.  These limitations have motivated a number of 
quasielastic scattering experiments employing hadronic 
probes\cite{lampfpp,triumfpp,lampfpn,hadron} .  
The more complicated interaction of the probe 
with nucleons means that -- in principle -- new responses which are 
inaccessible with electrons can be studied.  One important example 
of such investigations is found in the $(\vec p,\vec p')$ (Refs.
\cite{lampfpp,triumfpp}) and $(\vec p,\vec n)$ (Refs.
\cite{lampfpn})
experiments performed (in part) to extract the 
{\it longitudinal} spin response to complement information on the
transverse spin response determined by electron scattering.  Analysis of
such hadronic measurements is, however, hampered by the strength and 
complexity of the projectile- nucleon interaction.  The former causes
strong distortions in the incoming and outgoing projectile wavefunctions
and localizes the scattering process in the region of the nuclear surface.
The latter typically implies significant modification of the interaction 
in the nuclear medium.  Both sets of effects greatly complicate the
theoretical description of the scattering process and the extraction
of nuclear structure information.

	These considerations have led to much interest in $K^+$-nucleus
scattering.  The relative weakness of the $K^+N$ interaction
relative to, {\it e.g.}, $NN$ and $\pi N$ interactions suggests that 
distortions and medium effects may be simpler to handle and that the
scattering process may be more sensitive to the nuclear interior.  
Estimates of the mean-free-paths (MFP's) of the various probes support
the latter assertion.  For example, at a laboratory momentum of 
700 MeV/c, the proton and the $\pi^+$ have MFP's of about 2 fm while
the $K^+$ MFP is roughly twice that value.

	While $K^+$-nucleus scattering experiments are still in their 
infancy, their promise has been some what diminished by early results.
In particular, there is a persistant discrepancy between measured 
$K^+$-nucleus elastic and total cross sections and theoretical results
based on multiple scattering models which are expected to be accurate
due again to the weakness of the $K^+N$ interaction\cite{elastic}.

	Data for $K^+$-nucleus quasielastic scattering from C, Ca and
Pb at a laboratory $K^+$ momentum of 705 MeV/c have recently become
available\cite{prl,prc} .  A preliminary theoretical analysis of the
$|\bfq|$= 300 and 500 MeV/c data for $^{12}$C and $^{40}$Ca has already 
appeared along with the data\cite{prl}.  In the present paper, 
we extend those preliminary
calculations and focus on comparison with $^{40}{\rm Ca}(e,e')$ data 
at $|\qs|=500$ MeV/c.  In so doing, we make the first attempt 
to realize the promise of $K^+$ scattering to provide nuclear structure
information complementary to that extracted from electron data.  The 
remainder of this paper is organized as follows: in Section II, we
derive well known results for $(e,e')$ quasielastic scattering so
as to establish a framework for treating $K^+$ scattering which
is addressed in Section III.  Section IV touches upon problems 
associated with ambiguities in the forms of both the $eN$ and 
$K^+N$ on-shell elastic scattering amplitudes.  Past efforts to 
resolve the $eN$ ambiguity are summarized and a resolution for  
$K^+N$ is proposed.  The theoretical treatment of nuclear structure 
is outlined in Section V.  Specific calculations of electron and $K^+$ 
cross sections and comparisons with data appear in Section VI where
it is shown that, due to details of the structure of the $K^+$ 
quasielastic cross section, the strong quenching observed for the
$(e,e')$ Coulomb response does {\it not} appear for the $K^+$ 
process even though the latter is dominated by a contribution which
is superficially very similar to the Coulomb response.  Section VII 
contains a summary of the present work and our conclusions.
 
\section{FORMALISM FOR ELECTRON-NUCLEUS SCATTERING}

	The discussion of electron scattering from nucleons and nuclei
presented in this section is considerably more detailed than would 
seem warranted given that no new results appear.  However, this 
detail is supplied so as to provide a familiar context for developing our 
treatment of $K^+$ scattering which will be outlined in Section III.

	The unpolarized $e^-$ scattering differential cross section in 
plane wave Born approximation is
\begin{equation}
	d\sigma =
	{1\over{v_{rel}}}\ 
	{{4\alpha^2}\over{(q^2)^2}}\ \ell_{\mu\nu}\ W^{\mu\nu}_{EM}\
	d^3 p_f  \label{ba}
\end{equation}
where $q=p_i-p_f=(\omega,\bfq)$ and $p_i=(\epsilon_i,\bfp_i)$, 
$p_f=(\epsilon_f,\bfp_f)$ are the initial and final electron 
4-momenta, respectively, and the electron electromagnetic tensor is
\begin{eqnarray}
	\ell_{\mu\nu}&=&{1\over 2}\sum_{s_f s_i}\ {\rm Tr}\biggl[ 
	\gamma_\mu\ u(p_f s_f)\bar u(p_f s_f)\ \gamma_\nu\ 
	u(p_i s_i)\bar u(p_i s_i)\biggr] \nonumber\\ 
	&=&{1\over{\epsilon_f\epsilon_i}}\ 
	\bigl[p_\mu p_\nu + {1\over 4}
	(q^2 g_{\mu\nu}-q_\mu q_\nu)\bigr] \label{bb}
\end{eqnarray}
where $p={1\over 2}(p_i+p_f)$ and, {\it e.g.},
\begin{equation}
	u(p_i,s_i)=\sqrt{{{\epsilon_i + m}\over{2\epsilon_i}}}
	\left(\matrix{ 1 \cr
		       {{\bfsigma\cdot\bfp_i}\over{\epsilon_i+m}}
			\cr}\right) \chi(s_i)  
\end{equation}
where $m$ is the electron mass.
Also, 
\begin{equation}
	W^{\mu\nu}_{EM}=-{1\over \pi}\ {\rm Im}\ 
	\Pi^{\mu\nu}_{EM}(\bfq,\bfq;\omega) \label{bj}
\end{equation}
is the electromagnetic response of the scatterer.  For nuclear scattering, 
\begin{equation}
	\Pi^{\mu\nu}_{EM}(\bfq,\bfq';\omega)\equiv
	\int d^3y\ e^{-i\bfq\cdot\bfy}\ 
	\int d^3y'\ e^{+i\bfq'\cdot\bfy'}\
	\Pi^{\mu\nu}_{EM}(\bfy,\bfy';\omega) \label{bk}
\end{equation}
where the polarization insertion $\Pi^{\mu\nu}_{EM}(\bfy,\bfy';\omega)$
is defined via
\begin{eqnarray}
	\Pi^{\mu\nu}_{EM}(y,y')&\equiv&
	{1\over i}
	<i|T[\hat{\bar\psi}(y)\Gamma^\mu_{EM}\hat\psi(y)\ 
	\hat{\bar\psi}(y')\Gamma^\nu_{EM}\hat\psi(y')]|i> \nonumber\\
	&=&\int{{d\omega}\over{2\pi}}\ e^{-i\omega(y_0-{y_0}')}\ 
	\Pi^{\mu\nu}_{EM}(\bfy,\bfy';\omega). \label{bl}
\end{eqnarray}
In these expressions, the nucleon electromagnetic current operator is
\begin{equation}
	\Gamma^\mu_{EM}\equiv F_1(q^2)\gamma^\mu 
	+ i F_2(q^2){{\sigma^{\mu\nu}(P_f-P_i)_\nu}\over{2M}}  \label{bm}
\end{equation}
where $M$ is the nucleon mass and $P_i$ ($P_f$) is the initial (final)
nucleon momentum.

	We now consider $e$-nucleon scattering.  It is useful to recall
the Lehmann representation of the polarization insertion which implies, 
{\it e.g.}, 
\begin{equation}
	-{1\over \pi}\ {\rm Im}\ 
	\Pi^{\mu\nu}_{EM}(\bfq,\bfq;\omega)=\sum_f\ 
	\bigl(J^\mu_{fi}\bigr)^*\  J^\nu_{fi} 
	\quad \delta (\omega-E_f+E_i) 
\end{equation}
where 
\begin{equation}
	J^\mu_{fi}=\int d^3y\ e^{+i\bfq\cdot\bfy}\ 
	<f|\hat{\bar\psi}(\bfy)\ \Gamma^\mu_{EM}\ \hat\psi(\bfy)|i> .
	\label{bi}
\end{equation}
For a single nucleon, $J^\mu_{fi}$ is readily evaluated and we find
\begin{equation}
	W^{\mu\nu}_{EM}\rightarrow\delta(\omega-E_f+E_i)\ 
	\delta_{\bfq, {\bfP}_f-{\bfP}_i}\ \ell^{\mu\nu}(N)
\end{equation}
where the Kroenecker $\delta$ comes from the box normalization of the 
free nucleon wavefunction and where the {\it nucleon} electromagnetic
tensor is
\begin{eqnarray}
	\ell^{\mu\nu}(N)&\equiv&{1\over 2}\sum_{S_f S_i}\ {\rm Tr}\ 
	\biggl[
	\Gamma^\mu_{EM}\ u(P_f S_f)\bar u(P_f S_f)\ \Gamma^\nu_{EM}\ 
	u(P_i S_i)\bar u(P_i S_i)\biggr] \nonumber\\
	&\rightarrow&{1\over{E_f E_i}}\biggl[
	{{G^2_E+\tau G^2_M}\over{1+\tau}}\ P^\mu P^\nu
	+G^2_M\ {1\over 4}\bigr(q^2 g^{\mu\nu}-q^\mu q^\nu\bigl)\biggr]
	\label{bo}
\end{eqnarray}
with $P\equiv{1\over 2}(P_f+P_i)$, $\tau\equiv -q^2/4M^2=Q^2/4M^2$ 
and Sach's form factors, $G_M=F_1 + F_2$ and $G_E=F_1 -\tau F_2$.

	Then, in the initial nucleon rest frame, using $\wp\equiv|\vec p|$,
\begin{equation}
	{{d\sigma}\over{d\Omega}}\bigg|_{lab}=
	\int^\infty_{0^+}{{\epsilon_i}\over{\wp_i}}\ 
	{{4\alpha}\over{(q^2)^2}}\ \ell_{\mu\nu}\  \ell^{\mu\nu}(N)\ 
	\delta(\epsilon_i-\epsilon_f+E_i-E_f)\ 
	\delta_{\bfp_i-\bfp_f,\bfP_f-\bfP_i}\ \ \wp^2_f d\wp_f .
\end{equation}
Evaluating the integral {\it at fixed scattering angle} 
yields
\begin{equation}
	{{d\sigma}\over{d\Omega}}\bigg|_{lab}=
	{{4\alpha}\over{(q^2)^2}}\ \ell_{\mu\nu}\  \ell^{\mu\nu}(N)\ 
	{{\epsilon_i}\over{\wp_i}}\ 
	{{\wp_f E_f}\over
	  {\bigg|\wp_f(\epsilon_i+M)
	   -\epsilon_f\wp_i\cos\theta\bigg|}}\ 
	\wp_f\epsilon_f \label{bg}
\end{equation}
where we now understand that $\wp_f$, $\epsilon_f$ and $E_f$ take on their
on-shell values in the final state.

	For an infinitely massive structureless proton, $\tau\rightarrow 0$, 
$E_f\rightarrow M$, $\wp_f=\wp_i$, $\epsilon_f=\epsilon_i$ and 
$F_1\rightarrow 1$, $F_2\rightarrow 0$ which implies $G^E$, 
$G^M\rightarrow 1$.  In this limit,
\begin{equation}
	\ell^{\mu\nu}(N)\rightarrow g^{\mu 0} g^{\nu 0}
\end{equation}
and
\begin{equation}
	q^2=-\bfq^2=-4\wp^2_i\sin\theta/2 .
\end{equation}
We then have
\begin{equation}
	{{d\sigma}\over{d\Omega}}\bigg|_{lab}\rightarrow\sigma_M= 
	{{\alpha^2}\over{4\beta^2\wp_i^2\sin^4\theta/2}}\ 
	(1-\beta^2\sin^2\theta/2)
	\label{bh}
\end{equation}
which is the Mott cross section.  For ultrarelativistic electrons, this
expression becomes
\begin{equation}
	\sigma_M\rightarrow\sigma_M'= 
	{{\alpha^2\ \cos^2\theta/2}\over{4\epsilon_i^2\sin^4\theta/2}} . 
\end{equation}
For physical nucleons; {\it i.e.}, when no approximation to $\ell^{\mu\nu}
(N)$ is made, we find, for ultrarelativistic electrons, the familiar result
\begin{equation}
	{{d\sigma}\over{d\Omega}}\bigg|_{lab}=
	\sigma_M'\ \biggl(1-{{2\tau M}\over{\epsilon_i}}\biggr)\ 
	\biggl[ {{G^2_E+\tau G^2_M}\over{1+\tau}} 
	+ 2\tau G^2_M\tan^2\theta/2\biggr] . \label{bbb}
\end{equation}

	Next, we examine $e$-nucleus scattering.  Using $\wp d\wp
=\epsilon d\epsilon$, Eq.~\ref{ba} yields, in the nuclear rest frame, 
\begin{equation}
	{{d^2\sigma}\over{d\omega d\Omega}}= 
	{{4\alpha^2}\over{(q^2)^2}}\  
	{{\epsilon_i\wp_f\epsilon_f}\over{\wp_i}}\ 
	\ell_{\mu\nu} W^{\mu\nu}_{EM}
	\label{bn}
\end{equation}
Using the fact that $q_\mu W^{\mu\nu}_{EM}=q_\nu W^{\mu\nu}_{EM}=0$
due to current conservation, we find
\begin{eqnarray}
	{{d^2\sigma}\over{d\omega d\Omega}}=
	{{4\alpha^2}\over{(q^2)^2}}\ {{\wp_f}\over{\wp_i}}\ m^2\ 
	&\biggl\{& 
	\biggl( {{\epsilon_f+\epsilon_i}\over{2m}}\biggr)^2\ 
	\biggl[\biggl({{Q^2}\over{\qs^2}}\biggr)^2 W^{00}_{EM}
	+ {{Q^2}\over{\qs^2}} W^{11}_{EM} \biggr] \nonumber\\
	&\qquad&\qquad\qquad -\biggl(1+{{Q^2}\over{4m^2}}\biggr) 
	W^{11}_{EM} \nonumber\\
	&\qquad&-{{Q^2}\over{4m^2}}\ {{Q^2}\over{\bfq^2}}\ W^{00}_{EM} +
	{{Q^2}\over{2m^2}}\ W^{11}_{EM}
	\ \ \biggr\} . \label{bc}
\end{eqnarray}
For ultrarelativistic electrons this reduces to the familiar 
expression\cite{familiar}
\begin{equation}
	{{d^2\sigma}\over{d\omega d\Omega}}= 
	\sigma_M'\ 
	\biggl[\biggl({{Q^2}\over{\qs^2}}\biggr)^2\ W^{00}_{EM}
	+\biggl({{Q^2}\over{2\qs^2}}+\tan^2\theta/2\biggr)\ 2W^{11}_{EM}
	\biggr] \label{bd}
\end{equation}
in which we identify the Coulomb and transverse responses, namely, 
$W_C=W^{00}_{EM}$ and $W_T=2W^{11}_{EM}$, respectively.
We now observe that the expression for the ultrarelativistic $eN$ cross 
section in the initial nucleon rest frame (Eq.~\ref{bbb}) can be expressed as
\begin{equation}
	{{d\sigma}\over{d\Omega}}(eN)=
	\int\ d\omega\ {{d^2\sigma}\over{d\omega d\Omega}}
	(eN)\bigg|_{\rm fixed\  angle}
\end{equation}
where
\begin{eqnarray}
	{{d^2\sigma}\over{d\omega d\Omega}}(eN)=
	\sigma_M'\ \delta(\omega&-&E_f+M)\ \delta_{\bfq,\bfP_f}\ 
	\nonumber\\ \times
	&\biggl[& \biggl({{Q^2}\over{\bfq^2}}\biggr)^2\ 
	{{1+\tau}\over{1+2\tau}}G^2_E
	+\biggl({{Q^2}\over{2\bfq^2}}+\tan^2\theta/2\biggr)\ 
	2\ {{\tau}\over{1+2\tau}}G^2_M \biggr] .
\end{eqnarray}
Comparison with Eq.~\ref{bd} allows us to determine the {\it single 
nucleon} electromagnetic responses
\begin{eqnarray}
	W^{00}_{EM}&\rightarrow&W^{00}_{EM}(N)=
	\delta(\omega-E_f+M)\ \delta_{\bfq,\bfP_f}\ {{1+\tau}\over{1+2\tau}}
	G^2_E \label{be} ,\\ 
	W^{11}_{EM}&\rightarrow&W^{11}_{EM}(N)=
	\delta(\omega-E_f+M)\ \delta_{\bfq,\bfP_f}\ {{\tau}\over{1+2\tau}}
	G^2_M . \label{bff}
\end{eqnarray} 
These expressions also follow directly from Eq.~\ref{bi} when applied to
a free nucleon.

\section{FORMALISM FOR KAON-NUCLEUS SCATTERING}

	This section presents a treatment of $K^+N$ and $K^+$-nucleus
scattering which emphasizes the relation to the formulation of $e^-$ 
scattering appearing in the previous section.  We begin by considering the 
(fictitious) scattering of a $K^+$ which interacts only electromagnetically
with the scatterer.  (Alternatively, we can think of the scattering of a 
``spinless electron''.)  Using arguments analogous to those which led to 
Eq.~\ref{ba} in Section II, we find that the $K^+$ {\it electromagnetic} 
cross section is given in plane wave Born approximation by
\begin{equation}
	d\sigma =
	{1\over{v_{rel}}}\ 
	{{4\alpha^2}\over{(q^2)^2}}\ L_{\mu\nu}\ W^{\mu\nu}_{EM}\
	d^3 p_f  \label{ca}
\end{equation}
where now $q=p_i-p_f$ where $p_i$ and $p_f$ are the initial and final 
$K^+$ momenta and $L_{\mu\nu}$ is the $K^+$ electromagnetic tensor
given by
\begin{eqnarray}
	L_{\mu\nu}&=&{1\over{4\epsilon_f\epsilon_i}}\ 
	(p_i+p_f)_\mu\ (p_i+p_f)_\nu \nonumber\\
	&=&{1\over{\epsilon_f\epsilon_i}}\ p_\mu\  p_\nu .
\end{eqnarray}
This expression is to be compared with the electron electromagnetic
tensor, Eq.~\ref{bb}. Differences are clearly due to the presence of 
``spin currents'' in the electron case.  

	We now treat $K^+N$ electromagnetic scattering.  It is simple
to show that, in analogy with Eq.~\ref{bg}, 
\begin{equation}
	{{d\sigma}\over{d\Omega}}(K^+N-EM)\bigg|_{lab}=
	{{4\alpha}\over{(q^2)^2}}\ L_{\mu\nu}\  \ell^{\mu\nu}(N)\ 
	{{\epsilon_i}\over{\wp_i}}\ 
	{{\wp_f E_f}\over
	  {\bigg|\wp_f(\epsilon_i+M)
	   -\epsilon_f\wp_i\cos\theta\bigg|}}\ 
	\wp_f\epsilon_f . \label{cb}
\end{equation}
For a massive structureless proton, we have
\begin{equation}
	{{d\sigma}\over{d\Omega}}(K^+N-EM)\bigg|_{lab}\rightarrow
	{{\alpha^2}\over{4\beta^2\wp_i^2\sin^4\theta/2}}\ 
\end{equation}
which should be compared with the Mott cross section appearing in 
Eq.~\ref{bh}.  (Since the limit of ultrarelativistic $K^+$'s is irrelevant 
for the measurements to be discussed below, we do not present the
corresponding formulae.)  The $K^+$ electromagnetic analogue to the
$e$-nucleus double differential cross section appearing in Eq.~\ref{bc} is
\begin{eqnarray}
	{{d^2\sigma}\over{d\omega d\Omega}}=
	{{4\alpha^2}\over{(q^2)^2}}\ {{\wp_f}\over{\wp_i}}\ m^2\ 
	\biggl\{ 
	&\biggl(& {{\epsilon_f+\epsilon_i}\over{2m}}\biggr)^2\ 
	\biggl[\biggl({{Q^2}\over{\qs^2}}\biggr)^2 W^{00}_{EM}
	+ {{Q^2}\over{\qs^2}} W^{11}_{EM} \biggr] \nonumber\\
	&\qquad&\qquad\qquad -\biggl(1+{{Q^2}\over{4m^2}}\biggr) 
	W^{11}_{EM} \ \ \biggr\} \label{cc}
\end{eqnarray}
where now $m$ is the $K^+$ mass.

	Our next task is to generalize this treatment to handle
{\it hadronic} interactions.  For {\it electromagnetic} $K^+$
scattering, the $t$-matrix which leads to Eq.~\ref{ca} is explicitly
\begin{equation}
	T_{fi}={1\over V}\ {1\over{\sqrt{4 \epsilon_f \epsilon_i}}}\ 
	{{-e^2}\over{q^2}}\ 
	p_\mu\ J^\mu_{fi}(\bfq) .
\end{equation}
where $V$ is the box volume arising from box normalizing the free
$K^+$ wavefunctions and where the electromagnetic transition current
$J^\mu_{fi}$ is defined in Eq.~\ref{bi}.  We next write  
\begin{eqnarray}
	{{-e^2}\over{q^2}}\ 
	p_\mu\ J^\mu_{fi}(\bfq)&\rightarrow&
	{\cal F}_{EM}'\ {{p_\mu}\over{m}}\ J^\mu_{fi}(EM),\quad
	{\cal F}_{EM}'\equiv{{-m e^2}\over{q^2}} . 
\end{eqnarray}
Clearly, the dynamics specific to electromagnetic scattering appear 
only in the quantity ${\cal F}_{EM}'$; the remaining factors are of a 
form appropriate to a general 4-vector current-current interaction.
We propose that for $K^+$ {\it hadronic} scattering, we should let
\begin{equation}
	{\cal F}_{EM}'\ {{p_\mu}\over{m}}\ J^\mu_{fi}(EM)\rightarrow
	{\cal F}_S'\ J^s_{fi} + 
	{\cal F}_V'\ {{p_\mu}\over{m}}\ J^\mu_{fi} 
\end{equation}
where now a scalar interaction appears in addition to the vector term.
In this expression, ${\cal F}_S'$ and ${\cal F}_V'$ are complex numbers
related to the $K^+N$ elastic scattering amplitude in a manner to be
established below and where the exact definitions of $J^s_{fi}$ and 
$J^\mu_{fi}$ will be given shortly.  It is plausible that an interaction  
of this form -- while, as will be discussed below, not unique -- 
is sufficiently general since we know that an on-shell spin 0-spin $1/2$
elastic scattering amplitude can be completely specified 
at a given kinematic point (including an irrelevant overall phase) by
two complex numbers.  It is now convenient to introduce the notation
\begin{equation}
	{\cal F}_S'\ J^s_{fi} + 
	{\cal F}_V'\ {{p_\mu}\over{m}}\ J^\mu_{fi}\rightarrow
	f_a'\ J^a_{fi}(\bfq)
\end{equation}
where
\begin{equation}
	f_a'\equiv\cases{ {\cal F}_S' \quad $ for $ a=s \cr
			  {\cal F}_V'\ {{p_\mu}\over{m}}$ for $ a=\mu \cr}
\end{equation}
and
\begin{equation}
	J^a_{fi}(\bfq)=\int d^3y\ e^{+i\bfq\cdot\bfy}\ 
	<f|\hat{\bar\psi}(\bfy)\ \Gamma^a\ \hat\psi(\bfy)|i>
\end{equation}
with
\begin{equation}
	\Gamma^a\equiv\cases{{\bf 1} $ for $ a=s \cr
			     \gamma^\mu $ for $ a=\mu \cr} . 
\end{equation}
With these definitions, it is straightforward to show that the 
{\it hadronic} $K^+$ cross section analogous to the electron and $K^+$ 
electromagnetic cross sections of Eqs.~\ref{ba} and \ref{ca}, respectively, 
is
\begin{equation}
	d\sigma |_{had} =
	{1\over{v_{rel}}}\ 
	{1\over{4\pi^2}}\ L_{ab}\ W^{ab}\
	d^3 p_f
\end{equation}
where the $K^+$ hadronic tensor $L_{ab}$ is defined as
\begin{equation}
	L_{ab}\equiv{{ {f_a'}^* f_b'}\over{4\epsilon_f\epsilon_i}}
\end{equation}
and the hadronic response is defined via (compare with Eqs.~\ref{bj} 
through \ref{bl})
\begin{eqnarray}
	W^{ab}&=&-{1\over \pi}\ {\rm Im}\ 
	\Pi^{ab}(\bfq,\bfq;\omega) \nonumber\\
	\Pi^{ab}(\bfq,\bfq';\omega)&=&
	\int d^3y\ e^{-i\bfq\cdot\bfy}\ 
	\int d^3y'\ e^{+i\bfq'\cdot\bfy'}\ 
	\Pi^{ab}(\bfy,\bfy';\omega)
\end{eqnarray}
where $\Pi^{ab}(\bfy,\bfy':\omega)$ is specified by
\begin{eqnarray}
	\Pi^{ab}(y,y')&=&{1\over i}
	<i|T[\hat\Gamma^a(y)\hat\Gamma^b(y')]|i> \nonumber\\
	&=&\int {{d\omega}\over{2\pi}}\ 
	e^{-i\omega(y^0-{y^0}')}\ 
	\Pi^{ab}(\bfy,\bfy';\omega)
\end{eqnarray}
and $\hat\Gamma^a(y)=\bar{\hat\psi}(y)\Gamma^a\hat\psi_(y)$.

	We now specifically consider $K^+N$ scattering.  In Eq.~\ref{bg}
for electron scattering and Eq.~\ref{cb} for $K^+N$ electromagnetic
scattering, we presented expressions for the differential cross sections 
{in the initial nucleon rest frame}.  It is straight forward to derive
a similar expression for the $K^+N$ {\it hadronic} cross section.  
A slight generalization yields this cross section in a reference frame in 
which the $K^+N$ relative motion is colinear but otherwise arbitrary.
It will be convenient to have a formula for the
$K^+N$ {\it hadronic} cross section {\it in the center-of-momentum frame}, 
namely
\begin{equation}
	{{d\bar\sigma}\over{d\Omega}}\biggl(K^+N\biggr)
	\bigg|_{cm}=
	{1\over 2}\sum_{S_f S_i}\bigg|f_a\ 
	\bar u(P_f S_f)\Gamma^a u(P_i S_i)\bigg|^2 \label{cd}
\end{equation}
where
\begin{equation}
	f_a\equiv {1\over{4\pi}}\ {E \over{\epsilon + E}}\ f_a' .
\end{equation}
and where $\epsilon$ and $E$ are the $K^+$ and $N$ energies, respectively, 
in the center-of-momentum frame.  It is straightforward to show
\begin{eqnarray}
	f_a\ \bar u(P_f S_f)\Gamma^a u(P_i S_i)&=&
	\bar u(P_f S_f)\biggl[ {\cal F}_S 
	+ {\cal F}_V {\pslash\over m}\biggr]u(P_i S_i) \nonumber\\
	&=&\chi^\dagger(S_f)\ \biggl[ F(q) + i\sigma_n G(q) \biggr]\ \chi(S_i)
	\label{cf}
\end{eqnarray}
where 
\begin{equation}
	{\cal F}_S={1\over{4\pi}}\ {E \over{\epsilon + E}}\ {\cal F}_S',
	\quad 
	{\cal F}_V={1\over{4\pi}}\ {E \over{\epsilon + E}}\ {\cal F}_V'
\end{equation}
and $\sigma_n\equiv\bfsigma\cdot\hat n$, $\hat n\equiv
(\bfp_i\times\bfp_f)/|\bfp_i\times\bfp_f|$ as well as 
\begin{eqnarray}
	F(q)={{E+M}\over{2E}}\biggl[ &{\cal F}_S&\biggl(
	     1-{{E-M}\over{E+M}}\cos\theta \biggr)
	    +{\epsilon\over m}{\cal F}_V\biggl(
	     1+{{E-M}\over{E+M}}\cos\theta \biggr)\biggr] \nonumber\\
	   &+&{{E^2-M^2}\over{2mE}}{\cal F}_V \biggl(1+\cos\theta\biggr)
\end{eqnarray}
and
\begin{equation}
	G(q)=\biggl\{ {{E-M}\over{2E}}\ \biggl[{\cal F}_S
	       -{\epsilon\over m}{\cal F}_V\biggr]
	     -{{E^2-M^2}\over{2mE}}{\cal F}_V \biggr\}\ \sin\theta . 
\end{equation}
We recognize $F(q)$ and $G(q)$ as the Wolfenstein amplitudes for 
$K^+$-nucleon scattering and take them from Arndt's SP88\cite{SAID} 
$K^+$-proton and $K^+$-neutron phase shift solutions.  The above relations 
may readily be inverted to find ${\cal F}_S$ and ${\cal F}_V$ or, 
equivalently, ${\cal F}_S'$ and ${\cal F}_V'$.

	We finally investigate $K^+$-nucleus scattering.  In analogy
with Eq.~\ref{bn} for $e$-nucleus scattering, the $K^+$-nucleus cross
section in the nuclear rest frame is
\begin{equation}
	{{d^2\sigma}\over{d\omega d\Omega}}\bigg|_{had}=
	{1\over{4\pi^2}}\ {{\epsilon_i\wp_f\epsilon_f}\over{\wp_i}}\ 
	L_{ab} W^{ab} . 
\end{equation}
Using $q_\mu W^{\mu\nu}=q_\nu W^{\mu\nu}=q_\mu W^{\mu s}=0$ which follows 
from the assumption that the baryon current is conserved and also using 
$W^{\mu s}=W^{s \mu}$, we can evaluate $L_{ab} W^{ab}$ to obtain -- 
in analogy with Eq.~\ref{bc} for $e$-nucleus scattering and  Eq.~\ref{cc} 
for $K^+$-nucleus electromagnetic scattering -- the following $K^+$-nucleus
hadronic cross section:
\begin{eqnarray}
	{{d^2\bar\sigma}\over{d\omega d\Omega}}\bigg|_{had}=
	{1\over{16\pi^2}}\ {{\wp_f}\over{\wp_i}}\ 
	&\biggl[&
	|{\cal F}_S'|^2 W^{ss} \nonumber\\
	&+&|{\cal F}_V'|^2 \biggl\{
	 \biggl({{\epsilon_f+\epsilon_i}\over{2m}}\biggr)^2\ 
	\biggl[\biggl({{Q^2}\over{\qs^2}}\biggr)^2 W^{00}
	+ {{Q^2}\over{\qs^2}} W^{11} \biggr] \nonumber\\
	&\qquad&\qquad\qquad -\biggl(1+{{Q^2}\over{4m^2}}\biggr) W^{11} 
	\biggr\} \nonumber\\
	&+&2{\rm Re}({\cal F}_S' {\cal F}_V'^*)\ 
	\biggl({{\epsilon_f+\epsilon_i}\over{2m}}\biggr)\ 
	{{Q^2}\over{\qs^2}} W^{0s}\ \ 
	\biggr] . \label{ce}
\end{eqnarray}
This is the main result of Section III.  We note that two new 
responses -- namely $W^{ss}$ and $W^{0s}$ -- which were not present for
the electromagnetic processes enter in the hadronic cross section.

	We finally observe that, for a single nucleon, the hadronic 
responses $W^{00}$ and  $W^{11}$ are given by the corresponding single 
nucleon electromagnetic responses
of Eqs.~\ref{be} and \ref{bff} in the limit that $F_1\rightarrow 1$ 
and $F_2\rightarrow 0$ or, equivalently, $G_E$, $G_M\rightarrow 1$.
Furthermore, it is easy to show that, in the initial
nucleon rest frame, $W^{ss}=W^{s0}=W^{00}=(1+\tau)/(1+2\tau)$ 
for a single nucleon.

\section{ON-SHELL AMBIGUITIES FOR PROJECTILE-NUCLEON INTERACTIONS}

	The expression for the electromagnetic current of a {\it free} 
nucleon (Eq.~\ref{bi}) contains the matrix element of the nucleon
electromagnetic current operator, $\Gamma^\mu_{EM}$ (Eq.~\ref{bm}),
evaluated between {\it free} nucleon Dirac spinors.  Using the 
Gordon decomposition, we may write
\begin{equation}
	\bar u(P_f,S_f)\ \Gamma^\mu_{EM}\ u(P_i,P_f)=
	\bar u(P_f,S_f)\ \Gamma^\mu_{EM}(\xi)\ u(P_i,P_f) \label{da}
\end{equation}
where
\begin{equation}
	\Gamma^\mu_{EM}(\xi)\equiv F_S(\xi){{(P_f+P_i)^\mu}\over
	{2M}}+F_V(\xi)\gamma^\mu+iF_T(\xi){{\sigma^{\mu\nu}
	(P_f-P_i)_\nu}\over{2M}} \label{daa}
\end{equation}
with
\begin{equation}
	F_S(\xi)\equiv -\xi F_2,\qquad F_V(\xi)\equiv F_1+\xi F_2,\qquad
	F_T(\xi)\equiv (1-\xi)F_2 \label{db}
\end{equation}
and where $\xi$ is {\it arbitrary}.  In reference to the {\it Dirac}
character of the operators in the three terms (as opposed to their
{\it Lorentz} character which is of course vectorial on all three cases),
we have identified the first through third terms as ``scalar'', ``vector'' 
and ``tensor'', respectively.  We may then transform to three distinct
on-shell equivalent forms of $\Gamma^\mu_{EM}$ each containing only two
terms according to
\begin{eqnarray}
	\xi=0\qquad&\leftrightarrow&\qquad{\rm VT}\nonumber\\
	\xi=1\qquad&\leftrightarrow&\qquad{\rm VS}\nonumber\\
	\xi=-F_1/F_2\qquad&\leftrightarrow&\qquad{\rm ST} \nonumber
\end{eqnarray}
where, {\it e.g.}, ``VT'' means that only vector and tensor terms are 
present.  (Note that, in evaluating the free nucleon electromagnetic 
tensor, $\ell^{\mu\nu}(N)$ of Eq.~\ref{bo}, it is most convenient to
employ the VS (or $\xi=1$) form of $\Gamma^\mu_{EM}(\xi)$.)

	The impulse approximation used in Section II to evaluate the
{\it nuclear} $(e,e')$ cross section does not in and of itself 
prescribe which representation of $\Gamma^\mu_{EM}(\xi)$ to use.  This
well-known ambiguity is problematic since as soon as the nucleon wave
functions are modified by the nuclear medium, the various forms of 
$\Gamma^\mu_{EM}(\xi)$ are no longer equivalent.  Put another way,
if we define the generalized electromagnetic response, 
$W^{\mu\nu}_{EM}(\xi)$, just as in Eqs.~\ref{bj} through \ref{bl}, 
except that $\Gamma^\mu_{EM}(\xi)$ is employed in Eq.~\ref{bl}, 
we find that $W^{\mu\nu}_{EM}(\xi)$ is not generally independent of
$\xi$.  This represents an important qualification of the oft-repeated 
statement that the nuclear electromagnetic interaction is well-understood.
(There are, of course, any number of other effects which can modify the 
in-medium electromagnetic interaction; we are at present restricting our
attention to the ambiguities inherent in the {\it impulse approximation}.)
It is worth noting that the $\xi$-dependence of electromagnetic observables
is especially strong in the relativistic model of nuclear structure we will
employ and which is outlined in Section V.  A definitive resolution of the 
$\xi$ ambiguity awaits a dynamical description of nucleon structure and how 
it is affected by the nuclear environment.  We do not, of course, propose
to solve that problem here.  Indeed, we simply adopt the VT or $\xi =0$ 
form of $\Gamma^\mu_{EM}(\xi)$ according to ``conventional 
wisdom''\cite{deforest}.

	Since the treatment of $K^+$-nucleus scattering presented in 
Section III is also based on the impulse approximation applied 
to the underlying
$K^+N$ interaction, it is plagued by its own on-shell ambiguities.  
In the expression for the center-of-momentum frame $K^+N$ cross section
(Eq.~\ref{cd}), we find the free nucleon matrix element
\begin{eqnarray}
	&&f_a\ \ \bar u(P_f S_f)\ \Gamma^a\ u(P_i S_i)\nonumber\\
	&=&\bar u(P_f S_f)\biggl[{\cal F}_S\ {\bf 1}+
	  {\cal F}_V{{\pslash}\over m}
	  \biggr]u(P_i S_i)\nonumber\\
	&=&\bar u(P_f S_f)\biggl\{ {\cal F}_S(\xi)\ {\bf 1}+{{p_\mu}\over m}
	  \biggl[{\cal F}_V(\xi)\gamma^\mu+i{\cal F}_T(\xi)
	  {{\sigma^{\mu\nu}(P_f-P_i)_\nu}\over{2M}}\biggr]\biggl\} 
	  u(P_i S_i) \label{dc}
\end{eqnarray}
where
\begin{equation}
	{\cal F}_S(\xi)\equiv {\cal F}_S
	+\xi {\cal F}_V{{p\cdot P}\over{mM}}\ ,\qquad
	{\cal F}_V(\xi)\equiv (1-\xi){\cal F}_V\ ,\qquad 
	{\cal F}_T(\xi)\equiv \xi {\cal F}_V  \label{dd}
\end{equation}
and $\xi$ is again arbitrary.
We have once more identified Dirac scalar, vector and tensor terms and may
also define scalar, vector and tensor forms of the interaction according to
\begin{eqnarray}
	\xi=0\qquad&\leftrightarrow&\qquad{\rm VT}\nonumber\\
	\xi=1\qquad&\leftrightarrow&\qquad{\rm VS}\nonumber\\
	\xi=-{\cal F}_S/\biggl({\cal F}_V{{p\cdot P}\over{mM}}\biggr)
	\qquad&\leftrightarrow&\qquad{\rm ST}. \nonumber
\end{eqnarray}
(We note that, by choosing the VT form of the $K^+N$ interaction, 
$K^+$ hadronic scattering becomes -- in plane wave Born approximation -- 
{\it formally} identical to the (fictitious) $K^+$ electromagnetic process
discussed in the beginning of Section III.)  As will be discussed in detail
below, $K^+$-nucleus cross sections show some sensitivity to the 
form of the $K^+N$ interaction.  Since there is no ``conventional wisdom'' 
to guide our choice as there was for electron scattering, we must appeal
to a dynamical model of $K^+N$ scattering for help in resolving the 
on-shell ambiguity.

	We turn specifically to the meson exchange model of $K^+N$ 
scattering developed by the Bonn group\cite{bonn} along the lines of their
model of the $NN$ interaction.  This model has as its input parameters 
meson masses, 
$K^+$-meson and $N$-meson coupling constants and form factors.  Certain
of the interactions implied by these couplings are iterated to all orders
and the resulting phase shifts are in good agreement with those 
determined by experiment. 
For our purposes, we focus on the $K^+N$ interactions
mediated by the exchange of $\rho$, $\omega$ and (fictitious) $\sigma$ 
mesons.  To proceed, we identify the T=0 (isoscalar) and T=1 (isovector)
combinations of the $K^+N$ elastic scattering amplitude.  In terms of,
{\it e.g.}, the Wolfenstein amplitudes of Eq.~\ref{cf}, we define
\begin{equation}
	F_{T=0}\equiv{1\over 2}\bigl(F_p+F_n\bigr)\ ,\qquad
	F_{T=1}\equiv{1\over 2}\bigl(F_p-F_n\bigr) \label{de}
\end{equation}
where $F_p$ ($F_n$) is the $K^+$-proton ($K^+$-neutron) amplitude.
The isoscalar and isovector pieces of ${\cal F}_S'$ and ${\cal F}_V'$ 
then follow
as described at the end of Section III.  For $N=Z$ targets, the $K^+$-nucleus
cross section is simply the sum of T=0 and T=1 contributions given by
Eq.~\ref{ce} with different ${\cal F}'$ 's and $W^{ab}$ 's for each isospin.

	In Born approximation, $\rho$ exchange contributes to the isovector 
amplitudes while $\omega$ and $\sigma$ exchange contribute to the
isoscalar amplitudes.  Since the $\rho N$ coupling has a vector-tensor
Dirac character (with the tensor component dominant), the {\it Born} 
isovector amplitude generated by $\rho$ exchange has a VT (or, referring 
to Eq.~\ref{db}, a 
$\xi=-{\cal F}_S/\biggl({\cal F}_V{{p\cdot P}\over{mM}}\biggr)$) 
form.  Numerically, the $\rho$ exchange Born amplitude predicted by the 
Bonn model at the kinematic point appropriate to the $|\bfq|=500$ MeV/c 
$K^+$-nucleus data, namely $\wp_i=705$ MeV/c and $\theta_{lab}=43$ 
degrees\cite{prl,prc}, is qualitatively consistent with the empirical 
amplitude\cite{SAID} we employ.  (Specifically, dominance of ${\cal F}_T$
over ${\cal F}_V$ is observed.)  Similarly, since the $\omega N$ coupling is 
assumed to be purely vector, the Born isoscalar amplitude due to $\sigma$
and $\omega$ exchange has a VS (or $\xi=0$) form.  Again, the Born 
amplitudes from the Bonn model are qualitatively consistent with the relevant
empirical results.  Specifically, it is observed that ${\cal F}_S$ and 
${\cal F}_V$ are of opposite signs (the scalar interaction is attractive) 
and are nearly equal in magnitude.

	The arguments presented above are clearly only suggestive.
While we have argued that the form for the $K^+N$ interaction employing the 
VS representation for the isoscalar amplitudes and the VT representation for 
the isovector amplitudes is physically most plausible, we will also show
in the analysis to follow 
calculations which utilize the VS representation for both the isoscalar and
isovector amplitudes.

\section{Nuclear Structure}  

In this section we describe our nuclear structure model.  Throughout this 
paper we assume the impulse approximation to be valid.  Hence, apart from
the ambiguities addressed in the previous section, 
the in-medium $e^{-}N$ and $K^{+}N$ couplings are 
entirely determined from on-shell data.  Moreover, we assume that both 
sets of couplings are small so that distortions and multi-step processes 
({\it e.g.}, multi-photon exchanges in the case of $(e,e')$) 
may be ignored. In this framework, all relevant 
nuclear structure information is contained in the responses of the nuclear 
target introduced in Sections II and III to be discussed further in this 
section. 

The response of the nuclear target will be calculated in a relativistic 
random-phase-approximation to the Walecka model. In the Walecka model
nucleons interact via the exchange of isoscalar scalar ($\sigma$) and 
vector ($\omega$) fields. The dynamics of the system is, thus, described 
in terms of the following Lagrangian density
\begin{eqnarray}
  {\cal L} &=&
  \bar{\psi}  ( i {\rlap/\partial} - M ) \psi             +
  {1 \over 2} ( \partial_{\mu}\phi\partial^{\mu}\phi
		    -m_{\rm s}^{2}\phi^{2} )              -
  {1 \over 4} F_{\mu\nu}F^{\mu\nu}  \nonumber \\         &+&
  g_{\rm s}\bar{\psi}\psi\phi                             -
  g_{\rm v}\bar{\psi}\gamma^{\mu}\psi V_{\mu}             + 
  \delta{\cal L} \;,
 \label{qhdI}
\end {eqnarray}
where $g_{\rm s}$($g_{\rm v}$) is the scalar(vector) coupling
constant, $M$, $m_{s}$, and $m_{v}$ are the nucleon, $\sigma$-meson, 
and $\omega$-meson masses, respectively, and $\psi$, $\phi$, and $V^{\mu}$ 
are the corresponding field operators. The term $\delta{\cal L}$ contains
renormalization counterterms and the antisymmetric field-strength tensor, 
$F_{\mu\nu}$, has been defined by
\begin{equation}
  F_{\mu\nu} \equiv \partial_{\mu}V_{\nu} - \partial_{\nu}V_{\mu} \;.
\end{equation}
The nuclear ground state will be obtained in a relativistic mean field
approximation to the Walecka model. In this case, the meson-field 
operators are replaced by their ground-state expectation values. This 
approximation yields a set of Dirac single-particle states that are 
determined self-consistently from the equations of motion.

The one-body response of the nuclear ground state to an external probe is 
fully contained in the polarization tensor. The polarization tensor 
is a fundamental many-body operator that can be systematically computed 
using well-known many-body techniques (e.g., Feynman diagrams).
To illustrate these techniques we concentrate on the vector polarization,
for simplicity. This is defined as the ground-state expectation value
of a time-ordered product of nuclear (vector) currents
\begin{equation}
  \Pi^{\mu\nu}(x,y) ={1\over i} 
    \langle i | T \Big[ J^{\mu}(x) 
    J^{\nu}(y) \Big] | i \rangle \;.
 \label{pimunu}
\end{equation}
In a mean-field approximation to the ground state the polarization
insertion can be written, exclusively, in terms of the single-nucleon 
propagator $G(x,y)$
\begin{equation}
  \Pi^{\mu\nu}(x,y) = {1\over i} {\rm Tr} 
    \left[ \gamma^{\mu}G(x,y)\gamma^{\nu}G(y,x) \right] \;.
 \label{pixy}
\end{equation}
The nucleon propagator contains information about the interaction 
of the nucleon with the average mean field provided by the nuclear 
medium. Note that even if the interactions are ignored, such as in 
a Fermi-gas description, the propagator would still be 
different than its free-space value because of the existence
of a filled Fermi sea. This fact suggests the following 
decomposition of the nucleon propagator
\begin{eqnarray}
  &&
  G(x,y) = \int_{-\infty}^{\infty} {d\omega \over 2\pi}
	   e^{-i\omega(x^{0}-y^{0})}
	   G({\bf x},{\bf y};\omega) \;, \\
  &&
    G({\bf x},{\bf y};\omega)     =
    G_{F}({\bf x},{\bf y};\omega) +
    G_{D}({\bf x},{\bf y};\omega) \;.
 \label{gxy}
\end{eqnarray}
The Feynman part of the propagator, $G_{F}$, has the same analytic 
structure as the free propagator, namely, antiparticle poles above 
the real axis, particle poles below the real axis, and residues
proportional to the single-particle wave functions
\begin{equation}
    G_{F}({\bf x},{\bf y};\omega) = \sum_{\alpha}
     \left[
       {     {U}_{ \alpha}({\bf x}) 
	 \overline{U}_{\alpha}({\bf y}) \over
	\omega - E_{\alpha}^{(+)} + i\eta     }   +
       {     {V}_{ \alpha}({\bf x}) 
	 \overline{V}_{\alpha}({\bf y}) \over
	\omega + E_{\alpha}^{(-)} - i\eta     }   
     \right] \;.
 \label{gfeyn}
\end{equation}
The density-dependent part of the propagator, $G_{D}$, corrects
$G_{F}$ for the presence of a filled Fermi surface. Formally, 
one effects this correction by shifting the position of the
pole of every occupied state from below to above the real axis
\begin{eqnarray}
    G_{D}({\bf x},{\bf y};\omega) &=& \sum_{\alpha < {\rm F}}
     U_{ \alpha}({\bf x})\overline{U}_{\alpha}({\bf y}) 
     \left[
	{1 \over \omega - E_{\alpha}^{(+)} - i\eta} -
	{1 \over \omega - E_{\alpha}^{(+)} + i\eta} 
     \right] \\ &=&
     2 \pi i \sum_{\alpha < {\rm F}}
     \delta\Big(\omega-E_{\alpha}^{(+)}\Big)     
	       {U}_{\alpha}({\bf x})        
      \overline{U}_{\alpha}({\bf y})        \;.
 \label{gdens}
\end{eqnarray}

The decomposition of the nucleon propagator into Feynman and 
density-dependent contributions suggests an equivalent decomposition 
for the polarization insertion
\begin{eqnarray}
  &&
  \Pi^{\mu\nu}(x,y) = \int_{-\infty}^{\infty} {d\omega \over 2\pi}
	      e^{-i\omega(x^{0}-y^{0})}
	     \Pi^{\mu\nu}({\bf x},{\bf y};\omega) \;, \\
  &&
  \Pi^{\mu\nu}({\bf x},{\bf y};\omega)     =
  \Pi^{\mu\nu}_{F}({\bf x},{\bf y};\omega) +
  \Pi^{\mu\nu}_{D}({\bf x},{\bf y};\omega) \;.
 \label{pixyfd}
\end{eqnarray}
The Feynman part of the polarization, or vacuum polarization,
$\Pi_{F}$, describes the excitation of nucleon-antinucleon 
($N\overline{N}$) pairs
\begin{equation}
  \Pi_{F}^{\mu\nu}({\bf x},{\bf y};\omega) ={1\over i} 
  \int_{-\infty}^{\infty}{d\omega' \over 2\pi}
  {\rm Tr} \Big[ 
     \gamma^{\mu} G_{F}({\bf x},{\bf y};\omega+\omega') 
     \gamma^{\nu} G_{F}({\bf y},{\bf x};\omega') 
	   \Big] \;.
 \label{pifeynman}
\end{equation}
Note that this contribution diverges and must be renormalized. 
A lowest order calculation of the response, however, requires 
of only the imaginary part of the polarization insertion. In 
infinite nuclear matter, the threshold for pair production is 
at $q_{\mu}^{2}=4M^{* 2} > 0$ ($M^{*}$ is the effective 
nucleon mass in the medium). This timelike threshold lies far 
away from the spacelike region accessible in electron and kaon 
scattering. Thus, a lowest-order description of the process is 
not sensitive to vacuum polarization. 
However, $N\overline{N}$ excitations can be virtually produced. 
Hence, in a more sophisticated treatment of the response (e.g., 
RPA) the effective coupling of the nucleon to the probe can be 
modified by vacuum polarization. Indeed, it has been suggested
that virtual $N\overline{N}$ pairs play an important role in the 
quenching of the Coulomb sum\cite{handp}.

The density-dependent part of the polarization, $\Pi_{D}$, is
finite and can be organized in terms of three distinct contributions
\begin{equation}
  \Pi_{D }^{\mu\nu}({\bf x},{\bf y};\omega) =
  \Pi_{FD}^{\mu\nu}({\bf x},{\bf y};\omega) +
  \Pi_{DF}^{\mu\nu}({\bf x},{\bf y};\omega) +
  \Pi_{DD}^{\mu\nu}({\bf x},{\bf y};\omega) \;,
 \label{pidens}
\end{equation}
with each one of them of at least linear in $G_{D}$
\begin{eqnarray}
  \label{pidall}
  \Pi_{FD}^{\mu\nu}({\bf x},{\bf y};\omega) &=&
   \sum_{\alpha<F} 
   \overline{U}_{\alpha}({\bf x}) \gamma^{\mu}
    G_{F}\Big({\bf x},{\bf y};E_{\alpha}^{(+)}+\omega\Big) 
    \gamma^{\nu} U_{\alpha}({\bf y})  \;,  
   \label{pifd} \\
  \Pi_{DF}^{\mu\nu}({\bf x},{\bf y};\omega) &=&
   \sum_{\alpha<F} 
   \overline{U}_{\alpha}({\bf y}) \gamma^{\nu}
    G_{F}\Big({\bf y},{\bf x};E_{\alpha}^{(+)}-\omega\Big) 
    \gamma^{\mu} U_{\alpha}({\bf x})  \;,  
   \label{pidf} \\
  \Pi_{DD}^{\mu\nu}({\bf x},{\bf y};\omega) &=& 2\pi i
   \sum_{\alpha<F} \sum_{\alpha'<F} 
   \Big[
    \overline{U}_{\alpha}({\bf x})\gamma^{\mu}U_{\alpha'}({\bf x}) 
   \Big]
   \Big[
    \overline{U}_{\alpha'}({\bf y})\gamma^{\nu}U_{\alpha}({\bf y}) 
   \Big]
   \delta\Big(\omega+E_{\alpha}^{(+)}-E_{\alpha'}^{(+)}\Big) \;.
   \label{pidd}
\end{eqnarray}
The density-dependent part of the polarization describes the 
traditional excitation of particle-hole pairs. A spectral 
decomposition of the Feynman propagator, for $\Pi_{FD}$, 
is useful when discussing the spectral content of the 
polarization (the $\Pi_{DF}$ term, with the opposite 
time ordering, contains the same physical information as $\Pi_{FD}$) 
\begin{equation}
  \Pi_{FD}^{\mu\nu}({\bf x},{\bf y};\omega) =
   \sum_{\alpha<F,\beta} 
   \left[
   {\overline{U}_{\alpha}({\bf x})\gamma^{\mu}U_{\beta}({\bf x})  
    \overline{U}_{\beta}({\bf y})\gamma^{\nu}U_{\alpha}({\bf y}) \over
    \omega - \Big(E^{+}_{\beta}-E^{+}_{\alpha}\Big) + i\eta} +
   {\overline{U}_{\alpha}({\bf x})\gamma^{\mu}V_{\beta}({\bf x})  
    \overline{V}_{\beta}({\bf y})\gamma^{\nu}U_{\alpha}({\bf y}) \over
    \omega + \Big(E^{-}_{\beta}+E^{+}_{\alpha}\Big) - i\eta} 
   \right] \;.
 \label{pifdspect}
\end{equation}   
The first term in the sum represents the formation of a 
particle-hole pair after the system has absorbed (e.g., a photon 
carrying) energy $\omega$. Note, however, that some of these 
particle-hole transitions should be Pauli-blocked since the Feynman 
part of the propagator includes an unrestricted ($\beta$) sum over 
all single-particle states. The role of $\Pi_{DD}$ in the present 
formalism is, precisely, to enforce the Pauli principle. Note that 
since the Feynman part of the propagator will be evaluated nonspectrally, 
these Pauli-forbidden transition can not be simply removed by hand. 

The density-dependent part of the polarization contains, in addition 
to particle-hole pairs, a contribution that has no nonrelativistic 
counterpart. This contribution is contained in the second term of the 
sum and represents the Pauli blocking of $N\overline{N}$ excitations. 
Recall that the Feynman part of the polarization insertion represents 
the unconstrained excitation of $N\overline{N}$ pairs. At finite density, 
however, some of these excitations should be Pauli-blocked.

It is important to note that the inclusion of antinucleon degrees of 
freedom is not an unnecessary complication. If one is satisfied with 
computing the lowest order, or uncorrelated, nuclear response, then a 
``nucleons-only'' approximation is certainly justified. If, however, 
one wishes to examine the role of correlations by means of an RPA 
response, then one is forced to include antinucleon degrees of freedom 
in order to satisfy fundamental physical principles such as gauge 
invariance.

Traditionally, relativistic calculations of the nuclear response
have been carried out using two mean-field approximations
to the Walecka model. In the mean-field theory (MFT) the Feynman
contribution to the single-particle propagator is neglected from
the calculation of the nucleon self-energy. In contrast, one 
incorporates the effect from the (full) Dirac sea in the 
relativistic Hartree approximation (RHA). For a mean-field ground 
state obtained in the MFT approximation, it has been shown that the 
consistent linear response of the mean-field ground state is
obtained by neglecting the Feynman part of the polarization insertion. 
This consistency is reflected, for example, in the proper treatment of
spurious excitations associated with an overall translation of the
center of mass. Notice, however, that in the MFT approximation one 
retains the Pauli blocking of an ($N\bar{N}$) excitation that has not 
been included from the outset. It has recently been shown that this 
approximation leads to severe inconsistencies in the description of 
the effective $\omega$-meson mass in the nuclear medium\cite{jean}. 
Thus, in
this work we favor an RHA treatment of the scattering process in which 
vacuum loops are included in both the description of the ground state 
as well as in the linear response of the system. 

We start with a discussion of the density-dependent contribution
to the polarization. This contribution is finite and can be evaluated
exactly in the finite system. According to Eq.~(\ref{pidall}) we
must compute --- self-consistently --- all occupied 
single-particle states and the Feynman part of the propagator.
The calculation proceeds by, first, calculating a set of occupied
single-particle states satisfying the following Dirac equation
\begin{equation}
 \Big[ E_{\alpha}^{(+)}\gamma^{0} + i{\bf \gamma}\cdot{\bf \partial}
      -M - \Sigma_{H}({\bf x}) \Big]U_{\alpha}({\bf x}) = 0 \;.
 \label{speq}
\end{equation}
A Hartree calculation of the mean-field ground state yields, in
addition to the single-particle spectrum, the self-consistent
(scalar and vector) mean-fields used to generate the spectrum 
\begin{equation}
 \Sigma_{H}({\bf x}) = \Sigma_{\rm s}({\bf x}) +
		       \Sigma_{\rm v}({\bf x})\gamma^{0} \;.
 \label{meanf}
\end{equation}
Knowledge of the self-consistent mean fields now enables one to
compute --- nonspectrally --- the Feynman part of the nucleon
propagator by solving the equation
\begin{equation}
 \Big[\omega\gamma^{0} + i{\bf \gamma}\cdot{\bf \partial}
      -M - \Sigma_{H}({\bf x}) 
 \Big]G_{F}({\bf x},{\bf y};\omega) = \delta({\bf x}-{\bf y}) \;,
 \label{greenseq}
\end{equation}
with the appropriate boundary conditions. 

The evaluation of the polarization insertion, although still highly 
nontrivial, gets simplified for the case of a spherically symmetric 
ground state. In this case, one can classify the single-particle
states according to a generalized angular momentum $\kappa$
\begin{equation}
 U_{E \kappa m}({\bf x}) = {1 \over x} \left( 
      \begin{array}{c}
       \phantom{i}g_{E\kappa}(x)
	{\cal Y}_{     \kappa  m}(\hat{\bf x})  \\
		 if_{E\kappa}(x)
	{\cal Y}_{\bar{\kappa} m}(\hat{\bf x})  
      \end{array} \right) \;,
\end{equation}
where $\bar{\kappa} \equiv -\kappa$ and we have introduced the
spin-spherical harmonics defined by
\begin{equation}
  {\cal Y}_{\kappa m}(\hat{\bf x}) = 
  \langle \hat{\bf x} | l {1 \over 2} jm \rangle \;, \quad
  \kappa = \cases{    l, & if $l=j+1/2\;;$ \cr
		   -l-1, & if $l=j-1/2\;.$ \cr }
 \label{curlyy}
\end{equation}
The Feynman part of the propagator can be, similarly, written
as a sum over partial waves
\begin{eqnarray}
 G_{F}({\bf x},{\bf y};\omega) = {1 \over xy} \sum_{\kappa m}
 \left(
  \begin{array}{rr}
    g_{11}^{\kappa}(x,y;\omega)
    {\cal Y}_{\kappa m}({\bf x})
    {\cal Y}^{\dagger}_{\kappa m}({\bf y}) &
  -ig_{12}^{\kappa}(x,y;\omega) 
    {\cal Y}_{\kappa m}({\bf x})
    {\cal Y}^{\dagger}_{\bar{\kappa} m}({\bf y}) \\
  +ig_{21}^{\kappa}(x,y;\omega) 
    {\cal Y}_{\bar{\kappa} m}({\bf x})
    {\cal Y}^{\dagger}_{\kappa m}({\bf y}) &
    g_{22}^{\kappa}(x,y;\omega) 
    {\cal Y}_{\bar{\kappa} m}({\bf x})
    {\cal Y}^{\dagger}_{\bar{\kappa} m}({\bf y}) 
   \end{array}
  \right) \;.
 \label{greensum}
\end{eqnarray}
Once the bound-state orbitals and the Feynman propagator have 
been determined, the evaluation of the polarization tensor in
momentum space 
\begin{equation}
  \Pi^{\mu\nu}({\bf q},{\bf q}';\omega) =
  \int d^{3}{x} \; e^{-i{\bf q}  \cdot {\bf x}}
  \int d^{3}{y} \; e^{ i{\bf q}' \cdot {\bf y}}
  \Pi^{\mu\nu}({\bf x},{\bf y};\omega) \;,
\end{equation}
becomes straightforward. The angular integrals are done 
analytically leaving two radial integrals to be performed 
numerically. We stress that the procedure outlined above
enables one to calculate the density-dependent part of the 
polarization exactly in the finite system.

The Feynman part of the polarization, however, must be
calculated in a local-density approximation. To our knowledge,
the renormalization of the divergent integrals has never been 
carried out in the finite system. Thus, in the present work we 
adopt the following form for the Feynman contribution to the
response
\begin{equation}
  {\Pi}_{F}^{\mu\nu}({\bf q},{\bf q}';\omega) =
  \int d^{3}{r} \; e^{-i({\bf q}-{\bf q}')\cdot{\bf r}}
  {\Pi}_{F}^{\mu\nu} 
   \Big(\bar{\bf q},\omega;M^{*}(r)\Big) \;,  
 \label{pilda}
\end{equation}
where ${\Pi}_{F}^{\mu\nu}\Big(\bar{\bf q},\omega;M^{*}(r)\Big)$
is the renormalized vacuum polarization calculated in nuclear matter at 
an average momentum $\bar{\bf q}=({\bf q}+{\bf q}')/2$, and at a local 
value of the effective nucleon mass 
\begin{equation}
  M^{*}(r) = M + \Sigma_{\rm s}(r) \;.
\end{equation}

The nuclear response will be calculated in a variety
of models and approximations. The most sophisticated calculation 
that we will present involves calculating the nuclear response in 
a relativistic random phase approximation (RPA) to the Walecka model.
In RPA one incorporates many-body correlations through an infinite
summation of the lowest order (or uncorrelated) polarization. Due
to scalar-vector mixing the RPA equations form a set of 
$5\times5$ coupled integral equations
\begin{equation}
	   {\Pi}_{\rm RPA}^{ab}({\bf q},{\bf q}';\omega) =
	   {\Pi}^{ab}({\bf q},{\bf q}';\omega) +
	   \int {d^{3}{k} \over (2\pi)^{3}}
	   {\Pi}^{ac}({\bf q},{\bf k};\omega) 
	    V_{cd}({\bf k};\omega)
	   {\Pi}_{\rm RPA}^{db}({\bf k},{\bf q}';\omega) \;,
 \label{rpaeq}
\end{equation}
where we have introduced latin indices $a\equiv(s,\mu)$
that run over scalar and vector Lorentz structures, and
a residual interaction $V_{ab}$ given by
\begin{equation}
  V_{ab}({\bf q};\omega) \equiv V_{ab}(q) =
  \left(
   \begin{array}{cc}
     g_{\rm s}^{2} \Delta(q)  &  0          \\
	  0  & g_{\rm v}^{2} D_{\mu\nu}(q) 
   \end{array}
  \right) \;.
\end{equation}
Note that the free vector and scalar propagators have been 
defined, respectively, by
\begin{eqnarray}
  D_{\mu\nu}(q) &=& 
   \left(-g_{\mu\nu}+q_{\mu}q_{\nu}/m_{\rm v}^2 \right)D(q) \;, \\
	D(q) &=& {1 \over q_{\mu}^{2} - m_{\rm v}^{2}} \;, \\
   \Delta(q) &=& {1 \over q_{\mu}^{2} - m_{\rm s}^{2}} \;.
\end{eqnarray}
The RPA equations are solved  --- for every spin and 
parity $J^{\pi}$ --- by, first, performing the radial 
($k$)-integral using a Gauss quadrature scheme and then
solving the resulting matrix equation using standard
matrix-inversion techniques.

We conclude this section with a brief discussion of the
response of infinite nuclear matter. Due to the 
translational-invariant character of nuclear matter
the previous discussion simplifies considerably.
In a mean-field approximation to the Walecka model
the meson-field operators are replaced by their classical
ground-state expectation values which are constants in 
nuclear matter
\begin{eqnarray}
  \phi    &\rightarrow& \langle \phi \rangle \equiv \phi_{0} \;, \\
  V^{\mu} &\rightarrow& \langle V^{\mu} \rangle 
	   \equiv g^{\mu 0}V^{0} \;.
 \label{mfields}
\end{eqnarray}
The ground-state of the system is, thus, characterized by
a filled Fermi (and Dirac) sea of nucleons with an effective
mass $M^{*}$ determined self-consistently from the 
equations of motion
\begin{equation}
  M^{*} \equiv M -  g_{\rm s}\phi_{0} \;,
\end{equation}
and effective nucleon and antinucleon energies which are
shifted by the presence of a constant vector field
\begin{eqnarray}
  && E^{(\pm)}_{\bf k} \equiv E^{*}_{\bf k} 
      \pm g_{\rm v}V^{0} \;, \\
  && E^{*}_{\bf k} \equiv \sqrt{{\bf k}^{2}+M^{* 2}} \;.
\end{eqnarray}
In particular, this implies that the nucleon propagator is, 
formally, indistinguishable from the free nucleon propagator.
The Feynman and density-dependent propagators, which are the 
basic building blocks for the response, are thus given, respectively, 
by  
\begin{eqnarray}
  G_{F}(k) &=& \Big( {\rlap/\bar{k}} + M^{*} \Big)
   \left[ {1 \over \bar{k}^{2} - M^{* 2} + i\eta} \right] \;, \\
 \label{gpf}
  G_{D}(k) &=& \Big( {\rlap/\bar{k}} + M^{*} \Big)
   \left[ 
     {i\pi \over E^{*}_{\bf k}}
     \delta\Big(\bar{k}^{0}-E^{*}_{\bf k}\Big)     
     \theta\Big({\rm k}_{\rm F}-|{\bf k}|\Big)  
   \right] \;,
 \label{gpd}
\end{eqnarray}
where $k_{\rm F}$ is the Fermi momentum and we have defined
\begin{equation}
  \bar{k}^{\mu} \equiv (k^{0}-g_{\rm v}V^{0}, {\bf k}) \;.
\end{equation}
From the nucleon propagator it is simple to construct the 
lowest-order nuclear response
\begin{equation}
 i\Pi^{\mu\nu}(q) = \int {d^{4}k \over (2\pi)^{4}}
  {\rm Tr} \Big[\gamma^{\mu}G(k+q)\gamma^{\nu}G(k)\Big] \;.
 \label{pinm}
\end{equation}
As before, the polarization contains a divergent Feynman component 
that must be renormalized, and a finite density-dependent contribution
that describes particle-hole excitations and the Pauli blocking 
of $N\overline{N}$ pairs. From this lowest order polarization
one computes the correlated response by solving the RPA equation 
which becomes, in nuclear matter, a simple algebraic equation.

The only ingredients that remain to be specified are the effective
number of nucleons and for nuclear-matter calculations the average
density at which the scattering occurs. These quantities are determined 
from eikonal formulae that read:
\begin{eqnarray} 
   A_{\rm eff} &=& \int d^{3}{r} \; e^{-\sigma t(b)} \rho(r) \;, \\
   \rho_{\rm eff}  &=& 
   {1 \over A_{\rm eff}}
   \int d^{3}{r} \; e^{-\sigma t(b)} \rho^{2}(r) 
   \equiv {2 k_{\rm F}^{3} \over 3\pi^{2}} \;, 
 \label{rhoeef}
\end{eqnarray}
where $\sigma$ is the elementary projectile-nucleon total cross 
section and $t(b)$ is the nuclear-thickness function defined by
\begin{equation}
  t(b) = \int_{-\infty}^{\infty} dz \; \rho(r) \;.
 \label{thickness}
\end{equation}
From the effective density a self-consistent nucleon mass $M^{*}$ 
is determined~\cite{serwal86} which then serves as input for the 
calculation of the various nuclear responses per nucleon. These responses 
are subsequently scaled by the effective number of nucleons $A_{\rm eff}$ 
and then compared to experiment. In the case of $e^{-}$-nucleus scattering 
we ignore the small electromagnetic distortions (i.e., assume 
$\sigma \equiv 0$) and compute for ${}^{40}$Ca: 
$A_{\rm eff}=40$, $k_{\rm F}=1.13$~fm$^{-1}$, and $M^{*}/M=0.81$.
The equivalent expressions for $K^{+}$-nucleus scattering
(where the isospin-averaged total cross section is $\sigma=14.12$~mb) 
become: $A_{\rm eff}=16.06$, $k_{\rm F}=1.04$~fm$^{-1}$, 
and $M^{*}/M=0.84$.  (Note that we use {\it experimentally}
determined\cite{prl,prc} values of $A_{\rm eff}$ to normalize the 
$K^+$ quasielastic calculations to be presented below.  The reason
for the discrepancy between our eikonal estimates and the experimental 
value of $A_{\rm eff}\simeq 21$ is not understood at present.)

\section{Results}

	We begin our comparison of $e^-$ and $K^+$ quasielastic
scattering by presenting calculations of the longitudinal (or Coulomb) and 
transverse $(e,e')$ responses for $^{40}{\rm Ca}$ at $|\qs|
=500$ MeV/c in Figures \ref{figa} and \ref{figb}, respectively.  The
data are from Reference \cite{electdat}.  All calculations are based on the 
relativistic Hartree approximation (RHA) to QHD as described in the
previous section and thus include effects due to polarization of the 
nucleon sea at the one-loop level.  We show both nuclear matter (NM)
and finite nucleus (FN) calculations without (Har) and with (RPA) RPA
correlations.  Figure \ref{figa} clearly shows the dramatic quenching of
the longitudinal response due to the RPA which results in reasonable 
agreement with experiment.  As explained in, {\it e.g.}, Ref. \cite{handp},
this quenching is most readily interpreted as a screening of the nucleon
charge due to polarization of the nucleon sea.  The magnitude of the
quenching is comparable for NM and FN calculations but an acceptable
description of the shape of the measured response evidently requires
inclusion of finite nucleus effects.  Figure \ref{figb} compares
our calculations with the measured transverse response.  Since, as for all 
calculations reported here, we have included isoscalar correlations only, 
the differences between Har and RPA results are small for this 
predominantly isovector response.  Good agreement with experiment is 
found for the low-energy 
side of the quasielastic peak, especially for the finite nucleus 
calculations. The underestimation of transverse 
strength on the high-energy side of the peak, believed to be dominated 
by isobar formation and meson-exchange currents, is a common 
shortcoming of most ``one-nucleon'' models such as ours. 

	Comparable calculations are displayed with the $K^+$ 
data\cite{prl,prc} in
Figures \ref{figc} and \ref{figd} where we employ the VS and mixed 
VS (for T=0) and VT (for T=1) representations, respectively, for the 
$K^+N$ $t$-matrix as discussed in Section IV.  In Figure \ref{figc}
the calculations are normalized using $A_{eff}=21$ which is the mean value
extracted from experiment\cite{prl,prc}.  
The calculations appearing in 
Figure \ref{figd} use $A_{eff}=24$ which is the experimental upper limit.
The agreement between our most complete  calculations, namely 
those labelled FN(RPA), and the data is excellent although the VS 
calculations reproduce the overall magnitude of the measured cross sections 
slightly better than the mixed VS-VT results which slightly underestimate
the measurements.  We note that, while the simple mass $M$ Fermi-gas
result reproduces the observed {\it peak} $K^+$ quasielastic cross
section of $\approx 0.24$ mb/sr/MeV using the eikonal value of 
$A_{eff}=16$, its accounting of the {\it shape} of the cross section is 
poor.  In particular, the mass $M$ Fermi-gas cross section peaks at
$\omega\approx 85$ MeV and drops too rapidly at high $\omega$, vanishing
at $\omega\approx 210$ MeV.  For this reason, the mass $M$ Fermi-gas 
calculations must be considered inadequate regardless of normalization.
Returning to the full calculations, we see that 
two differences relative to the $(e,e')$ calculations
are immediately apparent.  First, NM and FN calculations are much more 
alike for $K^+$ than for $(e,e')$.  Second and more striking is the fact
that RPA effects are relatively small for the $K^+$ cross sections in contrast
to the large quenching they generate for the $(e,e')$ longitudinal
response.  This difference is even more surprizing since the $K^+$ cross 
section is {\it dominated} by ``longitudinal'' contributions ({\it i.e.},
by terms in Eq. \ref{ce} not involving $W^{11}$) which are closely
related to the $(e,e')$ longitudinal response.  We will show that this
difference arises from a subtle interplay of kinematic and relativistic 
nuclear structure effects.  In the course of this discussion, we will see
that the lack of strong quenching in $K^+$ quasielastic scattering is
quite remarkable since the additional ``longitudinal'' responses
appearing in Eq. \ref{ce}, namely $W^{ss}$ and $W^{0s}$, are quenched
{\it even more} than $W^{00}$ to which the $(e,e')$ longitudinal response
is most closely related.

	To understand the important physical contributions which determine
the $(e,e')$ longitudinal response and the $K^+$ quasielastic cross
section, we focus on Eq. \ref{ce}, our plane-wave expression for the
latter.  With this formula as our starting point, we can define
\begin{equation}
	\sigma_L\equiv\int\ d\omega 
	{{d^2\bar\sigma}\over{d\omega d\Omega}}\bigg|_{had}=
	{1\over{16\pi^2}}\ {{\wp_f}\over{\wp_i}}\ \hat W^{00}
	\biggl[|{\cal G}_{V}'|^2+ 2 {\rm Re}\ {\cal G}_{V}'\ 
	{\cal F}_S'^*\ R^{0s} + |{\cal F}_S'|^2\ R^{ss}\biggr]
	\label{ffa}
\end{equation}
where $\hat W^{00}\equiv\int d\omega\ W^{00}$, $R^{0s}\equiv
\hat W^{0s}/\hat W^{00}$, $R^{ss}\equiv \hat W^{ss}/\hat W^{00}$ and
\begin{equation}
	{\cal G}_{V}'\equiv {{\epsilon_i+\epsilon_f}\over{2m}}\ 
	{{Q^2}\over{\qs^2}}\  {\cal F}_V' .
\end{equation}
The ``$L$'' subscript indicates we have dropped all transverse
({\it i.e.}, $\propto W^{11}$) contributions, retaining only
``longitudinal'' terms.  We furthermore consider only isoscalar
($T=0$) contributions since it is here that strong RPA correlations appear.
These restrictions are intended to simplify the analysis by focussing
exclusively on the physics of greatest interest.  We recall that the
isoscalar Coulomb response is roughly half of the full response.
To assess the relative importance of the longitudinal isoscalar contribution 
to the $K^+$ cross sections, we observe that, for $K^+N$ scattering from a 
free nucleon, the longitudinal $T=0$ and $T=1$ contributions are
$0.999$ and $0.017$ mb/sr, respectively, while the corresponding 
transverse contributions are $0.120$ and $0.210$ mb/sr.
This shows that the longitudinal isoscalar contribution accounts for about 
two thirds of the total.  Thus the isoscalar longitudinal contributions to 
which we temporarily restrict our attention for the sake of simplicity 
are very significant components of the measured $e^-$ and $K^+$ quantities.

	Let us now consider the specific case of $K^+$ scattering
from a free nucleon of mass $M$ originally at rest.  Then Eq. \ref{ffa} 
can be expressed as 
\begin{equation}
	\sigma_L\rightarrow\sigma_L^{(0)}=\kappa_0\ 
	{{1+\tau_0}\over{1+2\tau_0}}\ |{\cal G}_V'\ +
	{\cal F}_S'|^2 \label{ffb}
\end{equation}
where we have used the fact (Eq. \ref{be}) that $W^{00}=W^{0s}=
W^{ss}=(1+\tau_0)/(1+2\tau_0)$ which also implies
$R^{0s}=R^{ss}=1$.  In this expression we have also defined 
$\kappa_0\equiv 1/{16\pi^2}\cdot {\wp_f}/{\wp_i}$ and 
$\tau_0\equiv Q^2/4M^2$ and have redefined ${\cal G}_V'$, all at
the particular kinematics of this specific case.

	We now modify this situation by letting $M\rightarrow M^*$.
Since the kinematics change, we define
\begin{eqnarray}
	\alpha_1&\equiv&\biggl( {{\epsilon_i+\epsilon_f^*}\over{2m}}\ 
	{{{Q^*}^2}\over{\qs^2}} \biggr)\ \div \ \biggl(
	{{\epsilon_i+\epsilon_f^{(0)}}\over{2m}}\ 
	{{{Q^{(0)}}^2}\over{\qs^2}} \biggr),  \nonumber\\
	\beta_1&\equiv& {{1+\tau^*}\over{1+2\tau^*}}\ \div \ 
	{{1+\tau_0}\over{1+2\tau_0}},  \nonumber\\
	\gamma&\equiv& \alpha_1^2\cdot \wp_f^*/\wp_f^{(0)}  \nonumber
\end{eqnarray}
and
\begin{equation}
	f(\alpha)\equiv {1\over{\alpha^2}}\ 
	{{|\alpha {\cal G}_V'\ + {\cal F}_S'|^2}
	\over{|{\cal G}_V'\ + {\cal F}_S'|^2}}
	\label{ffc}
\end{equation}
where now the ``starred'' quantities refer to the {\it new} kinematics
of the $M\rightarrow M^*$ case while the ``zero'' designation is for the 
mass $M$ kinematics.  We still have $R^{0s}=R^{ss}=1$ and can write
\begin{equation}
	{\sigma_L^{(0)}}^*={\sigma_L^{(0)}}\cdot 
	\gamma\beta_1\ f(\alpha_1) . \label{ffd}
\end{equation}

	Next we let $k_F\ne 0$ and write
\begin{equation}
	\hat W^{00}\rightarrow\beta_2\ {{1+\tau_0}\over{1+2\tau_0}},\quad
	\nonumber
\end{equation}
and
\begin{equation}
	R^{0s}\rightarrow\alpha_2^{-1} . \label{ffe}
\end{equation}
Then, because $R^{ss} = (R^{0s})^2$ is an excellent approximation, we have
\begin{equation}
	\sigma_L^*= {\sigma_L^{(0)}}\cdot 
	\gamma\beta_2\ f(\alpha_2) . \label{fff}
\end{equation}
where $\sigma_L^*$ is now equal to the integrated uncorrelated nuclear
matter cross section per nucleon obtained from the cross sections designated 
in Figures \ref{figc} and \ref{figd} by NM (Har).

	Finally we consider the effects of the RPA.  With
\begin{equation}
	\hat W^{00}\rightarrow
	\beta_3\cdot{{1+\tau_0}\over{1+2\tau_0}} \nonumber
\end{equation}
and
\begin{equation}
	R^{0s}\rightarrow \alpha_3^{-1}  \label{ffg}
\end{equation}
we find
\begin{equation}
	\sigma_L^{RPA}= {\sigma_L^{(0)}}\cdot 
	\gamma\beta_3\ f(\alpha_3) \label{ffh}
\end{equation}
where $\sigma_L^{RPA}$ is equal to the integrated uncorrelated nuclear
matter cross section per nucleon designated in Figures \ref{figc} and 
\ref{figd} by NM (RPA).
We also observe that the uncorrelated integrated longitudinal $(e,e')$
response per nucleon; {\it i.e.}, the uncorrelated Coulomb sum per nucleon, 
is given by
\begin{equation}
	{\cal S}_L^*=\beta_2\cdot \biggl(
	{{1+\tau_0}\over{1+2\tau_0}} \biggr) .
	\label{ffi}
\end{equation}
Its correlated counterpart is
\begin{equation}
	{\cal S}_L^{RPA}=\beta_3\cdot \biggl(
	{{1+\tau_0}\over{1+2\tau_0}} \biggr) .
	\label{ffj}
\end{equation}
The point of this formulation is that, by examining the numerical 
values of $\gamma$, $\alpha_i$ and $\beta_i$ as well
as the behavior of $f(\alpha)$, 
we can understand the physical relationships between the $K^+$ cross 
section and the longitudinal $(e,e')$ response.

	In pursuit of this understanding we first determine that
$\gamma=0.85$ for $\wp_i=700$ MeV/c and $M^*/M=0.81$ which is the 
self-consistent nuclear matter value for the average density of 
$^{40}$Ca.  (Note that, because of $K^+$ absorption in the nuclear 
medium, we should use the slightly larger value of $M^*/M$ appropriate
to the lower average density at which the $K^+N$ interaction occurs
for quasielastic scattering.  However, differences are small and
we use the overall average density of $^{40}$Ca, which is appropriate 
for electron scattering, in both cases so as to facilitate the following 
comparisons.)  It remains to establish the behavior of the $\beta_i$
and the $\alpha_i$ --- or, more relevantly, the values of $f(\alpha_i)$
--- as we include the effects of ({\it i}) $M^*\neq M$, ({\it ii})
$k_{F}\neq 0$ and ({\it iii}) RPA correlations.  These behaviors are
summarized in Figure \ref{fige} where we plot $\beta_i$, $f(\alpha_i)$
and $\sigma^{(i)}_L/\sigma^{(0)}_L=\gamma\beta_i f(\alpha_i)$ as they
evolve from case ({\it i}) through case ({\it iii}).  We see that
$M^*\neq M$ causes only a very small change in $\beta$.  Hence, the
ratio of the corresponding Coulomb sums, ${{\cal S}^{(0)}_L}^*/
{\cal S}^{(0)}_L=\beta_1$, is nearly unity which implies that changing
$M^*$ from $M$, by itself, has very little consequence for electron
scattering.  The effect on the integrated $K^+$ quasielastic cross
section is appreciable, however; we find $f(\alpha_1)=0.87$.  This 
large effect
is traceable to the delicate cancellation between the first two and the 
third terms of Eq. \ref{ce} and the changes in the kinematic factors 
multiplying these terms brought about by $M^*\neq M$.  Overall, taking 
into account the effect of $\gamma$, the $K^+$ cross section is reduced
by $0.73$ simply by letting $M^*/M\rightarrow 0.81$.

	We next examine the effect of $k_{F}\neq 0$.  The
$\beta$ factor does not change as the Coulomb strength is merely
redistributed.  However, now $R^{0s}\neq 1$ due to the Lorentz contraction
of the scalar density and in consequence the delicate cancellation 
alluded to above is somewhat altered.  We find $f(\alpha_2)=0.93$ and
the uncorrelated $K^+$ cross section goes up slightly so that 
$\sigma^*_L/\sigma^{(0)}_L=0.78$.

	We finally consider the influence of the RPA correlations.  As
shown in Figure \ref{fige}, they cause $\beta$ to drop dramatically to 
$0.61$!  This just reflects the strong RPA quenching of the 
Coulomb response which, as mentioned above, is a well-known feature of
the RHA RPA\cite{handp}.  We further observe that $R^{0s}$ 
goes from $0.978$ 
without correlations to $0.84$ with them which means that the summed
responses $\hat W^{0s}$ and $\hat W^{ss}$ are quenched even more strongly 
by the RPA than is $\hat W^{00}$!  Indeed, $\hat W^{ss}$ is reduced by
a factor of $0.46$.  This would seem to imply a {\it strong} quenching
of the $K^+$ cross section.  However, as is evident in Figure \ref{fige}, 
such is not the case because $f(\alpha_3)=1.38$, an increase which offsets 
the RPA quenching to the degree that $\sigma^{(i)}_L/\sigma^{(0)}_L$ drops
only from $0.78$ to $0.71$ ({\it i.e.}, by a factor of 0.91) when RPA
effects are included.  The large increase in $f(\alpha)$ in this case is 
due to the large reduction in $R^{0s}$, that is, due to {\it differential}
quenching of time-like vector and scalar contributions in the RPA.  If
no such differential quenching were present, $\sigma^{(i)}_L/\sigma^{(0)}_L$
would necessarily decrease like $\beta$;{\it i.e.}, like the Coulomb sum.
As the distinction between scalar and vector contributions is purely
relativistic, it is hard to see how non-relativistic models of nuclear
structure could simultaneously account for the strong quenching of the 
Coulomb sum and the absence of quenching in the $K^+$ quasielastic
scattering cross section in a manner as natural as for the present 
relativistic model.

	The preceeding analysis is based on a number of simplifying
assumptions which can be tested, {\it e.g.}, by comparing with the results
of more complete nuclear matter calculations which --- as we have already
established --- are quite consistent with the full finite nucleus results.
Figure \ref{fige} compares the ratio of the integrated nuclear matter
cross sections --- labelled ``NM'' --- with the $\sigma^{(i)}_L/
\sigma^{(0)}_L$ ratio discussed above.  The agreement is good enough
to inspire confidence in the simplified analysis.

	Overall, we find that the {\it full} integrated RHA-RPA $K^+$
quasielastic cross section per (effective) nucleon which includes
isoscalar transverse as well as isovector contributions {\it is}
reduced from the (isospin averaged) $K^+N$ cross section.  
For the VS representation of the $K^+N$ amplitude, the reduction factor
is 0.94 while for the mixed representation the factor is 0.82.  
These differences are clearly due to the different behavior of the 
isovector contributions, namely that they are appreciably enhanced by 
$M\rightarrow M^*$ in the VS representation but little changed in the VT
representation.
By comparison, the full integrated RHA-RPA Coulomb response per nucleon 
is reduced from the isospin averaged single nucleon value by a factor of 
0.76.  Clearly, while the $K^+$ reductions are less than for the Coulomb, 
they are not dramatically different, especially 
when the mixed representation is used for the former.  However, it is
important to observe that the reduction of the $K^+$ cross section 
is not due just to the quenching of the underlying responses but depends
also on the interference effects discussed above.  Most importantly, if
it were not for the {\it differential quenching} of scalar versus time-like
vector contributions which emerges naturally in our relativistic model
of nuclear structure, the reduction factor for the $K^+$ cross section 
would be {\it much} smaller and therefore inconsistent with the $K^+$ 
quasielastic data.  As it is, we have a gratifyingly accurate and 
consistent description of both the $K^+$ and $(e,e')$ data.

\section{Summary and Conclusions}

	We have formulated a treatment of $K^+$-nucleus quasielastic
scattering in a manner which parallels as closely as possible more-or-less 
standard treatments of $e^-$-nucleus quasielastic scattering.  The latter 
depends in a straightforward way on the Coulomb (or longitudinal) and 
transverse nuclear responses which in turn are of great importance in 
understanding essentials of nuclear structure.  We have shown that ---
in the present formulation --- $K^+$ quasielastic scattering depends on
these same Dirac vector and tensor nuclear responses as well as additional
ones containing Dirac {\it scalar} contributions.  Thus, in principle, 
$K^+$ quasielastic data can supplement and extend structure information 
extracted from the electron data and perhaps shed light on important issues 
such as the strong quenching of the Coulomb 
sum\cite{electdat,electdatb,coulthy}.  

	Our treatment of the underlying $K^+N$ interaction relies on the 
impulse approximation and we have been careful to spell out the
connection between the $K^+N$ amplitudes appearing in our expression for 
the $K^+$-nucleus quasielastic cross section, Eq.~\ref{ce}, and $K^+N$
phase shift solutions\cite{SAID}.  We have also briefly summarized problems 
associated with on-shell ambiguities in the form of the $e^-N$ amplitude
\cite{deforest} and have indicated how these problems carry over to the
form of the $K^+N$ amplitude.  We rely on a meson-exchange model of the 
$K^+N$ interaction\cite{bonn} to justify a specific form of this interaction
expressed as a ``mixture'' of Dirac vector and scalar invariants for the 
isoscalar channel and vector and tensor invariants for the isovector
channel.

	Our nuclear structure model is based on Quantum Hadrodynamics,
a successful relativistic phenomenology of nuclear dynamics.  We 
specifically focus on the Relativistic Hartree Approximation\cite{serwal86}
and the RPA based on it\cite{handp,toward}.  This treatment takes into 
account the polarization of the nucleon sea in one-loop approximation and
in so doing provides a unique mechanism for quenching the Coulomb 
response which is found to be in reasonable accord with experiment.
We have given a thorough discussion of both full finite nucleus and nuclear 
matter calculations of the nuclear responses in the RHA-RPA.  We also have
indicated how to fix the {\it effective} nuclear densities and $M^*/M$
values for the nuclear matter treatment of $K^+$ quasielastic scattering
which is complicated by the absorption of the $K^+$ scattering waves.

	We have compared our calculations with the Coulomb and transverse
responses for $^{40}$Ca at $|\qs|=500$ MeV/c\cite{electdat}.  We reproduce 
the low-$\omega$ side of the transverse response quite well, but, as is 
typical of ``one-nucleon'' models such as ours, we underestimate the
response on the high-$\omega$ side, presumably due to the omission of
meson-exchange-current and $\Delta$-isobar effects.  RPA effects strongly
quench the Coulomb response relative to the uncorrelated results and 
bring about reasonable agreement with the data.  This quenching can be
interpreted as a screening of the nucleon charge due to polarization of
the nucleon sea.  Finite nucleus effects appear to be important in 
reproducing the details of the shape of the measured Coulomb response.

	We have also shown similar calculations for the new 
$^{40}{\rm Ca}(K^+,{K^+}')$ data at the same momentum transfer
\cite{prl,prc}.  Here the quenching due to the RPA is much less than for 
the Coulomb response and differences between finite nucleus and nuclear 
matter calculations are smaller.  Agreement of the full RHA-RPA 
calculations with the measured cross sections is very good although
the calculations slightly underestimate the data when the ``mixed''
representation of the $K^+N$ amplitude is employed.

	We have gone on to explain why the RPA quenching of the $K^+$ cross
section is so much less than what is observed for the Coulomb response, a
phenomenon which is all the more surprizing given the dominanance of the
``longitudinal isoscalar'' contribution in the former which is where RPA
effects occur in our model.  The situation becomes even more puzzling upon
observing that the {\it new} responses containing scalar contributions 
which arise in the case of $K^+$ scattering (see Eq.~\ref{ce}) are even more 
strongly quenched than the Coulomb response.  However, careful analysis
shows that this {\it differential quenching} of responses alters a sensitive 
cancellation in the expression for the $K^+$ cross section in such a way that
the cross section is only slightly reduced.  The situation is qualitatively
unchanged for the full $K^+$ quasielastic cross section which also includes
transverse isoscalar and isovector components.  We note that the
phenomenon of differential quenching is purely relativistic in origin and
that, without it, the calculated $K^+$ cross sections would be much smaller 
and in strong disagreement with experiment.  We have concluded that our
relativistic model of nuclear structure provides a gratifyingly accurate
and consistent description of both $K^+$-nucleus and $e^-$-nucleus
quasielastic scattering.

\vskip 0.75 true in
 
\section{Acknowledgements} The authors gratefully acknowledge helpful
	comments by S.~Pollock and V.~Dmitrasinovi\' c.  
	This work supported in part by the U.S.D.O.E.

\vfill
\eject
\newpage
%
%
%
%

%
%
%
\newpage
\figure{ The longitudinal $(e,e')$ response, $S_L$,
	 for $^{40}{\rm Ca}$ at $|\qs|=500$ MeV/c.  Data are from
	 Ref.~\cite{electdat}.  
	 RHA nuclear matter (NM) and finite nucleus
	 (FN) calculations without (Har) and with (RPA) RPA  
	 correlations as descibed in the text are shown.
	 \label{figa}}
\figure{ Same as Figure \ref{figa}, but for the transverse $(e,e')$ 
	 response, $S_T$. \hfill
	 \label{figb}}
\figure{ The $K^+$ quasielastic cross section at
	 $p_K=703 {\rm MeV/c}$ and $\theta_{lab}=43$ degrees.
	 The data are from Refs.~\cite{prl,prc}.  RHA 
	 nuclear matter (NM) and finite nucleus
	 (FN) calculations without (Har) and with (RPA) RPA  
	 correlations and employing the VS representation of the $K^+N$
	 amplitude are shown.  The calculations are normalized
	 using $A_{eff}=21$, the centroid of the experimentally determined 
	 range of $A_{eff}$ for $^{40}$Ca \cite{prl,prc}.
	 \label{figc}}
\figure{ Same as Figure \ref{figc} except that the mixed representation
	 of the $K^+N$ amplitude is used and that the calculations are 
	 normalized using $A_{eff}=24$, the upper limit of the 
	 experimentally determined range of $A_{eff}$ for 
	 $^{40}$Ca \cite{prl,prc}.
	 \label{figd}}
\figure{ The evolution of the integrated longitudinal isoscalar $K^+$ 
	 cross section per nucleon is shown for ({\it i}) a single 
	 nucleon of mass $M$ initially at rest, ({\it ii}) a single 
	 nucleon of mass $M^*=0.81 M$
	 initially at rest, ({\it iii}) a Fermi gas of nucleons of mass
	 $M^*=0.81 M$ and Fermi momentum $k_{F}=1.13$ fm$^{-1}$ and 
	 ({\it iv}) a system like ({\it iii}) except with RPA correlations.
	 See discussion in text.
	 \label{fige}}

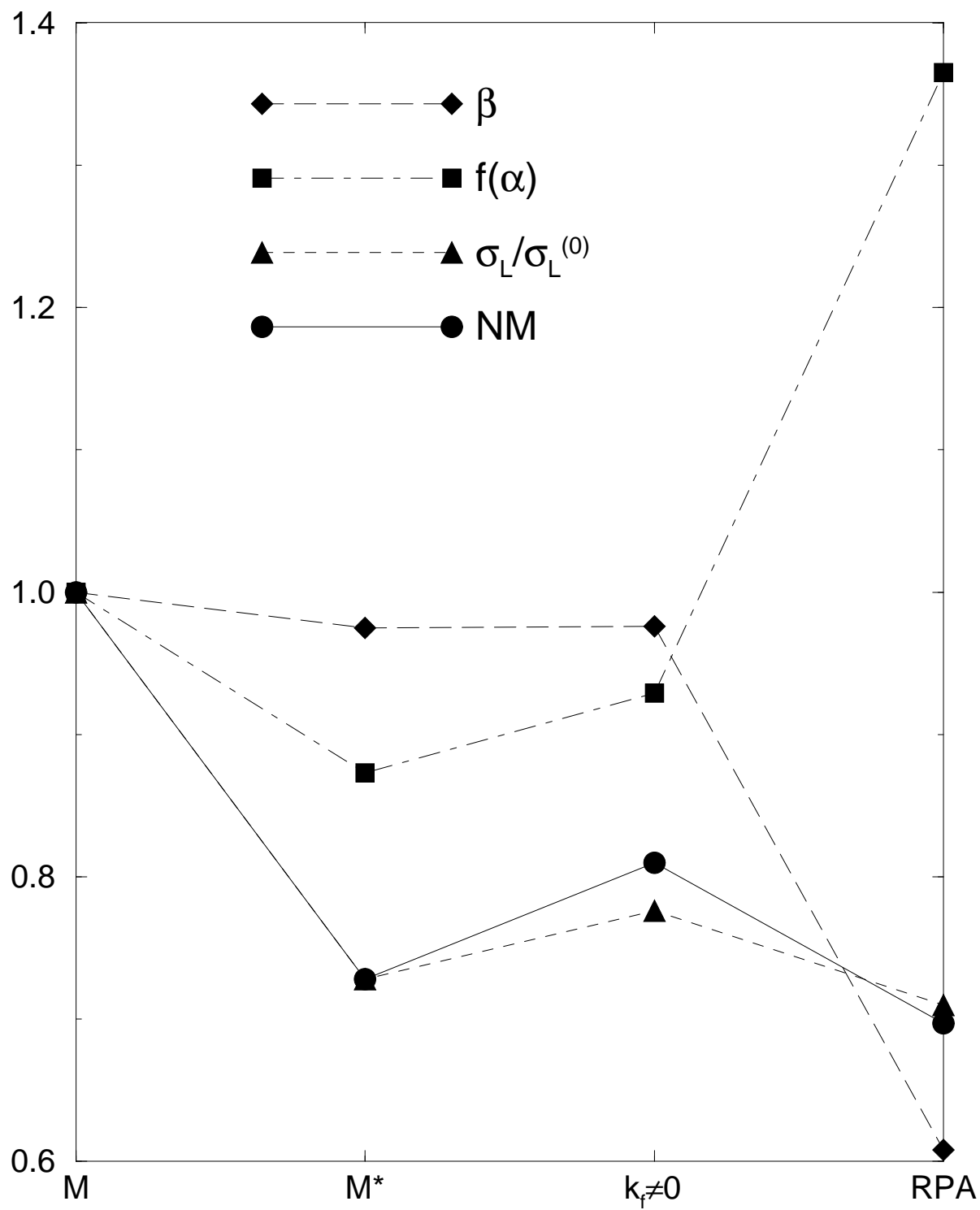
\end{document}